\documentclass[prd,aps,preprintnumbers,showpacs,10pt]{revtex4}
\usepackage{graphicx}
\usepackage{epstopdf}
\usepackage[centertags]{amsmath}
\usepackage{subfigure}
\usepackage{amsmath}
\usepackage{amssymb}
\usepackage{dsfont}
\usepackage{hyperref}

\begin{document}

\preprint{BARI-TH 633-2010}
\title{Finite temperature hadrons from holographic QCD}

\author{Floriana Giannuzzi}
\affiliation{Dipartimento di Fisica dell'Universit\`a degli Studi di Bari, I-70126 Bari, Italy\\  I.N.F.N., Sezione di Bari, I-70126 Bari, Italy}


\begin{abstract}
The properties of scalar mesons and glueballs at finite temperature are analyzed through a
{\it bottom-up} holographic approach. We focus on the spectral functions and mass
spectra. A discussion on hadron dissociation and deconfinement phase transition
is also put forward.
\end{abstract}

\maketitle

Since its appearence in 1998, the AdS/CFT correspondence  \cite{Maldacena:1997re} has been considered as a very promising tool for studying the non-perturbative regime of QCD, by relating it to a weakly-coupled theory. In particular, the correspondence can be better applied to the finite temperature case, one of the main reason being that in this limit the theory is no more conformal.

The conjecture states that type IIB string theory in a $AdS_5 \times S^5$ space, where $AdS_5$ is a five-dimensional anti-de Sitter space and $S^5$ is a five-dimensional sphere, is dual to $\mathcal{N}=4$ Super-Yang-Mills theory in a four-dimensional Minkowski space. It is also known as the holographic conjecture since the gauge theory can be constructed through a projection of the gravity theory on the boundary of the $AdS$ space \cite{Witten:1998qj}. 
According to the correspondence, a five-dimensional field $\phi$, whose boudary value is $\phi_0$, is related to a four-dimensional operator $\mathcal{O}$ by:
\begin{equation}\label{relAdSCFT}
Z_{\mathcal{S}}[\phi_0(x)]=\left< \mbox{e}^{ \int_{\partial AdS_{d+1}} \phi_0(x) \, \mathcal{O}(x)} \right>_{\mbox{\tiny CFT}} \,;
\end{equation}
strictly speaking, the generating functional on the left-hand side of (\ref{relAdSCFT}), computed in $\phi_0$, is equal to the generating functional of the correlation function of the operator whose source is $\phi_0$, on the right-hand side.  In Poincar\'e coordinates, the $AdS$ space is characterized by the metric
\begin{equation}
ds^2=\frac{R^2}{z^2}(dt^2-d\vec x^2-dz^2)\,, 
\end{equation}
where $z$ is called holographic coordinate. The boundary of the space is at $z=0$.

Up to now the dual theory of QCD has not been found yet. 
From a phenomenological point of view, people are trying to construct an {\it ad hoc} theory in a five-dimensional $AdS$ space such that its projection on the four-dimensional boundary can reproduce as many as possible QCD properties.
One of such {\it bottom-up} approaches developed so far is the Soft Wall model \cite{Karch:2006pv}, in which conformal symmetry (proper of $AdS$ spaces) is broken by inserting a factor $e^{-c^2\,z^2}$ in the action, with $c$ a mass scale, here fixed from the $\rho$ meson mass: $c=m_\rho/2=388$ MeV \cite{Karch:2006pv}.

In this holographic picture, temperature effects are introduced by modifying the metric of the anti-de Sitter space. In this respect, one can either impose a periodicity of Euclidean time, in which the temperature is the inverse of the compactification radius, or introduce a black hole in the metric along the fifth dimension, $z$, such that the temperature is related to the inverse of the position of the horizon of the black hole. From now on, the former case will be referred to as the ``Thermal-AdS'' model, whereas the latter case as ``AdS-Black Hole'' model. 
Therefore, ``Thermal-AdS'' model is characterized by  the following metric:
\begin{equation}\label{metricTH}
ds^2=\frac{R^2}{z^2} \left( d\tau^2+d\vec{x}^2+dz^2 \right) \qquad  0<\tau<\beta'=1/T\;, 
\end{equation}
where $\tau$ is the Euclidean time and $T$ is the temperature. On the other hand, the ``AdS-Black Hole'' metric is given by:
\begin{equation}\label{metricBH}
ds^2=\frac{R^2}{z^2} \left(f(z) d\tau^2+d\vec{x}^2+\frac{dz^2}{f(z)} \right) \qquad  f(z)=1-\frac{z^4}{z_h^4}\;, 
\end{equation}
where $z_h$ is the position of the horizon of the black hole, such that
\begin{equation}
 0<z<z_h=1/(\pi T) \;;
\end{equation}
in this case, the metric is smooth and complete if and only if the Euclidean time is periodic \cite{Witten:1998zw}, with period $\beta=1/T=\pi z_h$.\\
To find out what is the model that can better holographically describe QCD at finite temperature, one can either analyze both models separately and compare their outcomes with predictions coming from other approaches to QCD, or introduce a criterion for determining which metric is the stable one, for instance, by comparing the corresponding free energies and choosing the model with the lowest one. Here we consider both possibilities and compute spectral functions of scalar mesons and scalar glueballs.

\begin{figure}[h]
\includegraphics[width=8cm]{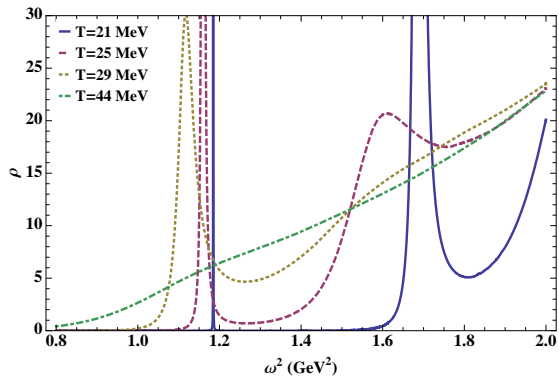}
\caption{Spectral function of scalar glueballs in the model with black hole (``AdS-Black Hole'') at $T=21$ MeV (solid blue line), $T=25$ MeV (dashed purple line), $T=29$ MeV (dotted yellow line), and $T=44$ MeV (dot-dashed green line).}
\label{specfunglu}
\end{figure}

\begin{figure}
\includegraphics[width=6.cm]{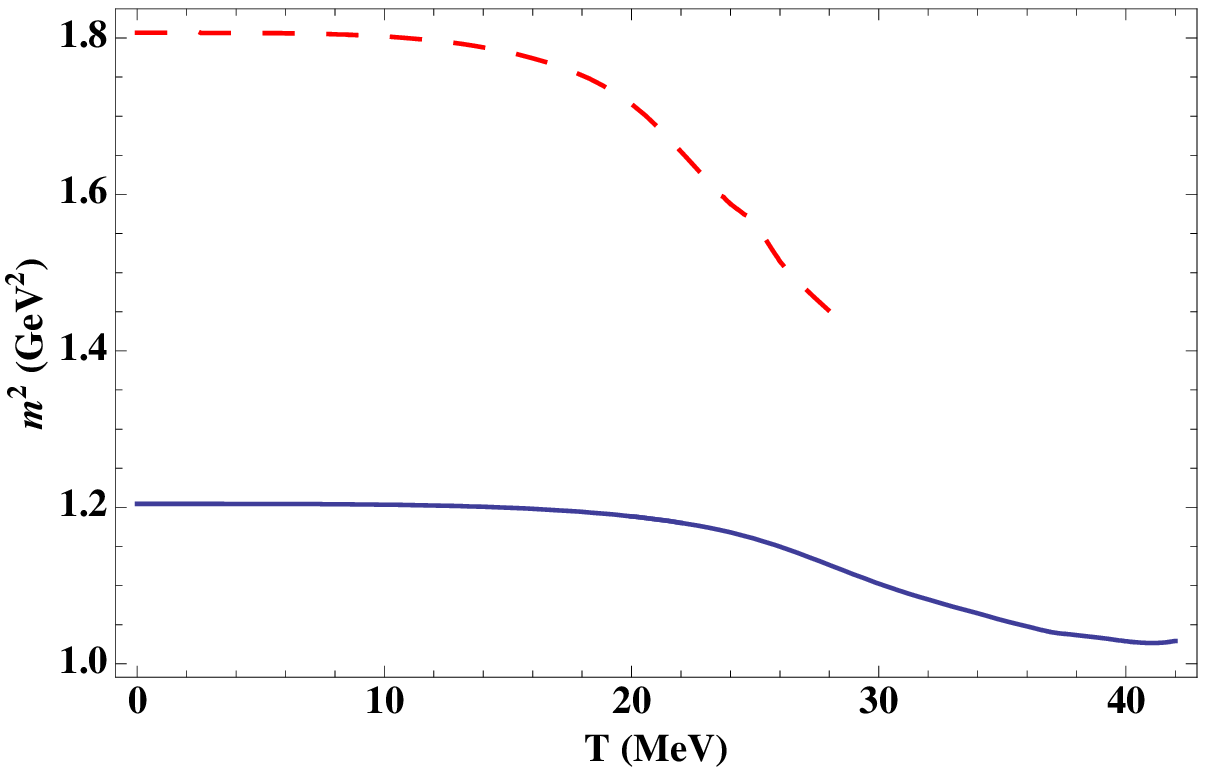}
\hspace*{.5cm}
\includegraphics[width=6.cm]{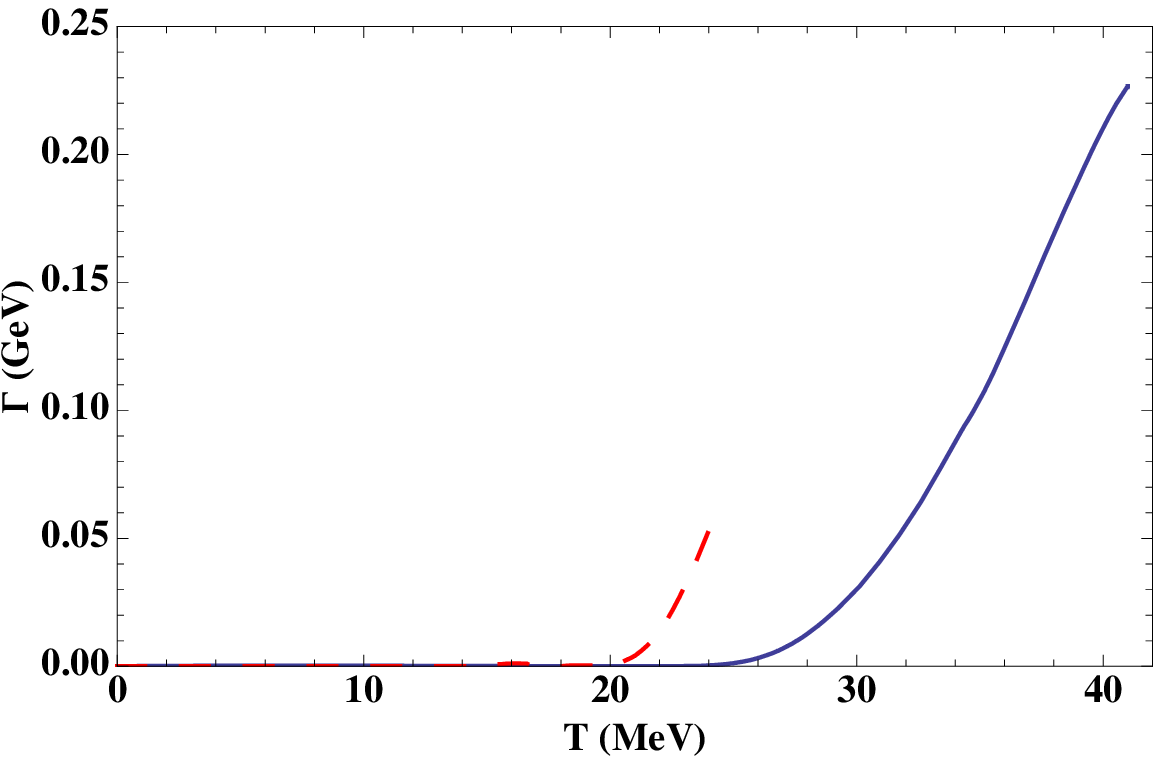}
\caption{Squared mass (left panel) and width (right panel) of scalar glueballs versus temperature in the model with black hole (``AdS-Black Hole'').}
\label{masswidglu}
\end{figure}

\begin{figure}
\includegraphics[width=6.cm]{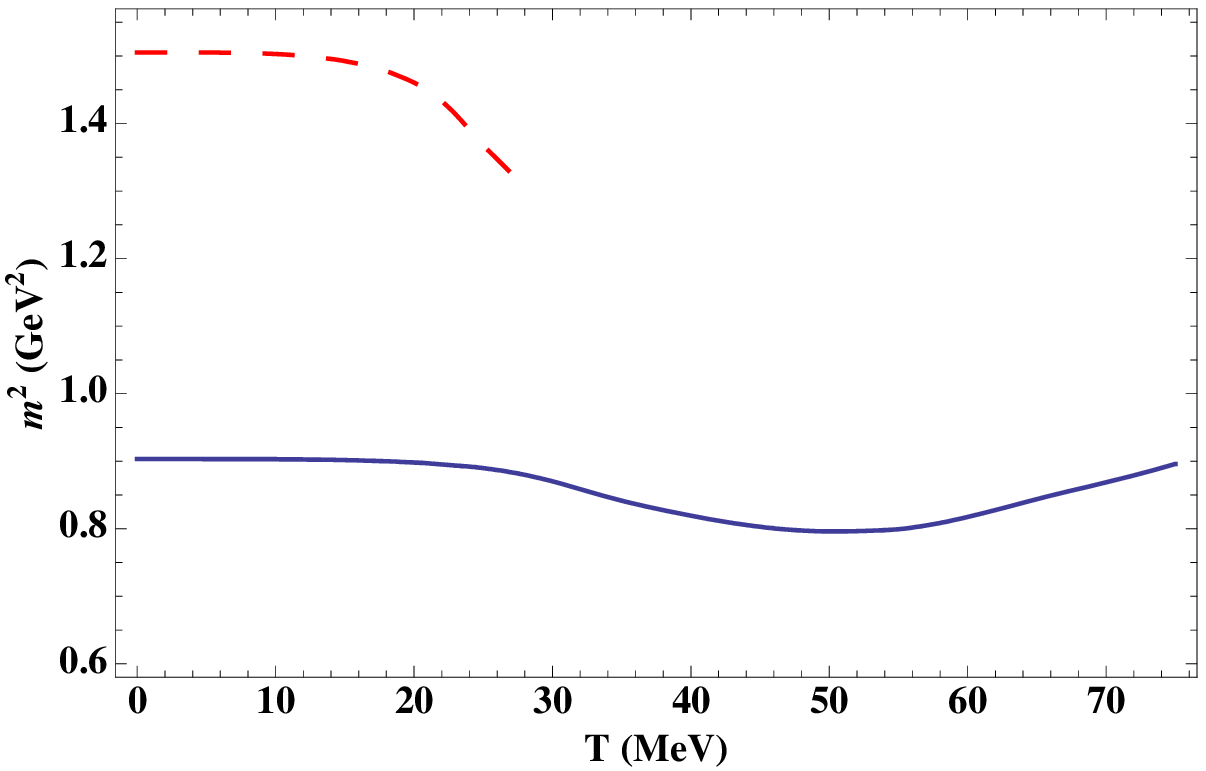}
\hspace*{.5cm}
\includegraphics[width=6.cm]{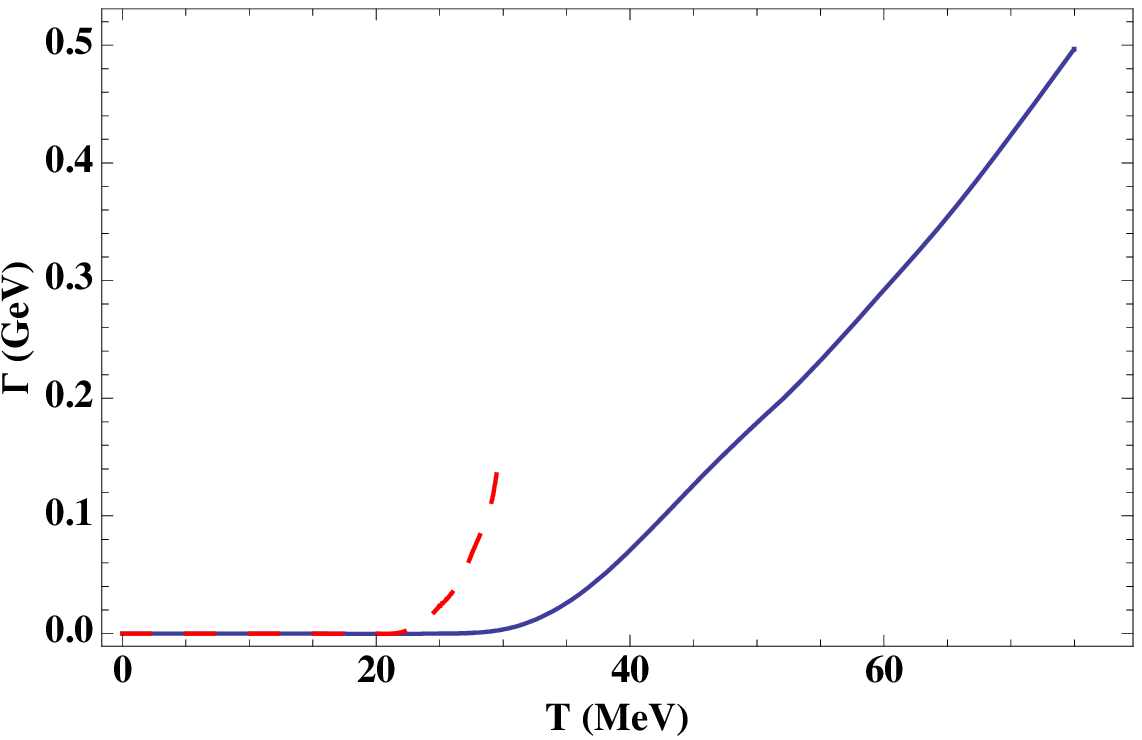}
\caption{Squared mass (left panel) and width (right panel) of scalar mesons versus temperature in the model with black hole (``AdS-Black Hole'').}
\label{masswidmes}
\end{figure}

Let us start from analyzing each model separately. From the point of view of spectral functions, ``Thermal-AdS'' is completely analogous to the zero-temperature model, and it yields the same results: finite-temperature spectral functions are therefore expected to be characterized by zero-width peaks at fixed positions,  as at $T=0$. \\
``AdS-Black Hole'', instead, deserves more attention. In the following, it will be shown how to compute spectral functions in a particular case, i.e. for scalar glueballs; however the procedure is general and can be extended to any other observable. Scalar glueballs have been investigated in the Soft Wall model at zero temperature in \cite{Colangelo:2007pt}. The five-dimensional field which is dual to the QCD operator $\beta(\alpha_s) \mbox{ Tr}(G^2(x))$, where $\beta(\alpha_s)$ is the Callan-Symanzik function, is a scalar massless field $X(x,z)$, standing the relation between the mass and the conformal dimension ($\Delta$) of a ($p$-form) operator \cite{Witten:1998qj}:
\begin{equation}
 m_5^2 R^2= (\Delta-p) (\Delta+p-4) \,.
\end{equation}
The five-dimensional action for this field in the Soft Wall model is:
\begin{equation}\label{actionglu}
\mathcal{S}=\frac{1}{2 k} \int d^5x\, \sqrt{g} \mbox{ e}^{-c^2\, z^2}\, g^{MN}\, \partial_M X  \partial_N X
\end{equation}
where $g$ is the determinant of the metric and $k$ is a parameter which makes the action dimensionless. In order to compute spectral functions we  move to the Fourier space by defining
\begin{equation}
X(x,z)=\int d^4q\,\, \mbox{e}^{iq\cdot x} \tilde{X}(q,z) \,,
\end{equation}
and we write the Fourier transformed field $\tilde{X}(q,z)=K(q,z) \tilde{X}_0(q)$ as the product of the bulk-to-boundary propagator $K(q,z)$ and the source $ \tilde{X}_0(q)$. From the action (\ref{actionglu}) one can derive the equation of motion for $K(q,z)$
\begin{equation}
K''(q,z)- \frac{4-f(z)+2\,c^2\,z^2\,f(z)}{z\, f(z)} K'(q,z)+\left( \frac{q_0^2}{f(z)^2}-\frac{\vec{q}^2}{f(z)} \right)  K(q,z)=0 \,;
\end{equation}
for semplicity, the case $\vec q=0$ (rest frame of the glueball) will be considered. The first boundary condition is $K(q,0)=1$, since the source $\tilde X_0$ is defined as the value of the field $\tilde X$ at $z=0$; the second one is that the bulk-to-boundary propagator must behave as a ``{\it falling in}'' solution near the horizon of the black hole:
\begin{equation}
K(q,u)\xrightarrow[u\rightarrow1]{}(1-u)^{-i \, q_0\, z_h/4}
\end{equation}
where $u=z/z_h$. The latter condition allows us to get in the end the retarded Green's function.
Standing Eq. (\ref{relAdSCFT}), the two-point correlation function can be computed by functionally deriving twice the action (\ref{actionglu}) with respect to the source $\tilde X_0(q)$, thus obtaining:
\begin{equation}
\Pi(q_0^2)=\left. \frac{\delta^2 \mathcal{S}}{\delta X_0 \delta X_0} \right|_{X_0=0}= \left. \frac{1}{2k}    \frac{R^3 f(u)}{u^3 z_h^4} \mbox{e}^{-c^2\, z^2} K(q,u) \partial_u K(q,u) \right|_{u=0}\,.
\end{equation}
The spectral function is the imaginary part of the Green's function $\rho(q_0^2)=\Im(\Pi(q_0^2))$; the first two peaks of the spectral function are shown in Fig. \ref{specfunglu} for four values of temperature \cite{Colangelo:2009ra}. The figure shows that at low temperatures the spectral function is characterized by very narrow peaks which become broader and move towards smaller values of mass as the temperature is increased; at $T\gtrsim 44$ MeV we can find no more peaks in the spectral function, so bound states do not exist anymore. We can also notice that excited states dissociate at lower temperatures than the ground state.
By fitting each peak through a Breit-Wigner function
\begin{equation}
\rho(x)=\frac{a\, m\, \Gamma\, x^b}{(x-m^2)^2+m^2 \Gamma^2} \,,
\end{equation}
it is possible to find how the squared masses and widths of glueballs vary with temperature; they are shown in Fig. \ref{masswidglu}. As expected, we find that the masses decrease while increasing temperature, starting from the value at $T=0$, $m^2_n=4 c^2(n+2)$, while the widths increase. 
The qualitative behavior we have found for spectral functions, masses and widths is similar to the one observed in lattice simulations \cite{lattice}, but the dissociation temperature for the ground-state glueball, i.e. the temperature at which the first peak in the spectral function disappears, turns out to be much lower than the one found in lattice studies.  \\
Similar results can be observed in the scalar-meson sector \cite{Colangelo:2009ra}; the two main differences are that the dissociation temperature for the ground state  (around 75 MeV) is higher than the one found for glueballs, and that the ground-state squared mass, although decreasing with temperature from its value at $T=0$ \cite{Colangelo:2008us}, at a certain temperature starts growing until the dissociation. The mass and width of scalar mesons versus temperature are plotted in Fig. \ref{masswidmes}. In general, one can see that the behavior of the spectral function in Fig. \ref{specfunglu} is quite universal, since it is similar for glueballs, scalar mesons and also vector mesons \cite{Fujita:2009wc}.

A further possibility for studying holographic QCD at finite temperature is to make the metric a dynamical quantity, so that it can vary with temperature. To determine which metric between ``AdS-Black Hole'' (\ref{metricBH}) and ``Thermal-AdS'' (\ref{metricTH}) should be used for each value of the temperature, one can compute and compare the corresponding free energies and choose the one with the lowest result \cite{Witten:1998zw}. In the Soft Wall model \cite{Herzog:2006ra}, the free energy in ``Thermal-AdS'' is
\begin{equation}
V_{TH}(\epsilon)=\frac{4R^3}{\kappa^2}\int_0^{\beta'} d\tau \int_\epsilon^{\infty} dz\,\, \frac{1}{z^5} \mbox{ e}^{-c^2\, z^2}\,,
\end{equation} 
and in ``AdS-Black Hole''
\begin{equation}
V_{BH}(\epsilon)=\frac{4R^3}{\kappa^2}\int_0^{\pi z_h} d\tau \int_\epsilon^{z_h} dz\,\, \frac{1}{z^5} \mbox{ e}^{-c^2\, z^2} \,,
\end{equation}
where $\epsilon\to 0$ has been introduced in order to regularize the two quantities. It turns out that at low ({\it resp.} high) temperatures ``Thermal-AdS'' ({\it resp.} ``AdS-Black Hole'') is the right metric to use. A first order Hawking-Page phase transition \cite{HPage} between the two metrics occurs at $T_c\approx 191$ MeV \cite{Herzog:2006ra}, and it has been identified with the deconfinement transition of QCD. 
As a matter of fact, in this framework the low-temperature spectral function has to be computed using the ``Thermal-AdS'' model, so it is characterized by zero-width peaks at fixed positions for every $T<T_c$. On the other hand, at temperatures  higher than the critical one Fig. \ref{specfunglu} shows that the spectral function computed in ``AdS-Black Hole'' has no peaks, so hadrons have already melted. Thus, in this model dissociation of scalar glueballs takes place together with deconfinement as a first order phase transition. This is also what happens to scalar mesons.

Some concluding remarks are then in order. We have analyzed two possible phenomenological models describing finite-temperature QCD in a holographic framework, looking at scalar-glueball spectral functions. If we use a model with a non-dynamical metric, in which the temperature is introduced through a black hole, we get a realistic qualitative description of the behavior of hadrons in a hot medium, but with a scale of temperature different from the one predicted by other models of QCD. If we let the metric change, describing the deconfinement transition on the boundary of the $AdS$ space as a Hawking-Page phase transition between two metrics in the bulk, the resulting spectral function in the confined phase is always equal to the one at zero temperature, while becoming suddenly flat in the deconfined phase. The first description can better simulate how hadron properties could change in a medium, but it seems to fail from a quantitative point of view. The second description may reproduce hadron properties in the limit of large $N_c$.
Therefore, the finite-temperature holographic representation of QCD in this {\it bottom up} approach still needs much efforts, and slight modifications of the model may be required. Recently, some developments of the ``AdS-Black Hole'' model have been put forward. In \cite{Colangelo:2010pe} the deconfinement transition in the chemical potential-temperature space is investigated in the Soft Wall model, using, as order parameter, the behavior of the static quark-antiquark potential at large distances. Furthermore, in \cite{Grigoryan:2010pj} the Authors construct a slightly different model, whose parameters are fitted from the masses and decay constants of $J/\psi$ and $\psi'$, thus finding higher dissociation temperatures also in the light-meson sector.

\vspace{0.5truecm}
 \textbf{Acknowledgments}
 \vspace{0.2truecm}

This work was supported, in part,  by the EU contract No. MRTN-CT-2006-035482, ``FLAVIAnet'' and by the grant ``\emph{Borse di ricerca in collaborazione internazionale}'' by \emph{Regione Puglia, Italy}. I thank the IPPP, Durham, for hospitality during the completion of this work.

Finally, I would like to join the Organizers of QCD@Work 2010 in warmly remembering Beppe Nardulli as a man devoted to science.

\end{document}